\documentclass[epj-spec]{svjour}
\usepackage{graphics}
\usepackage{epsfig}
\begin{document}
\title{Transmission of information and synchronization in a pair of coupled
chaotic circuits: an experimental overview}
\author{M. S. Baptista\inst{1} \thanks{\emph{Corresponding author:
baptista.murilo@gmail.com}} \and S. P. Garcia\inst{1}
\thanks{\emph{spinto@pks.mpg.de}; Present address: Signal Processing
Laboratory, IEETA, University of Aveiro, Campus Universit\'ario de
Santiago, 3810-193 Aveiro, Portugal.} \and S. K. Dana\inst{2}
\thanks{\emph{skdana@iicb.res.in}} \and J. Kurths\inst{3}
\thanks{\emph{juergen.kurths@pik-potsdam.de}}
}                     % Do not remove
\institute{Max-Planck-Institute f\"ur 
Physik komplexer Systeme, N\"othnitzer Str. 38, D-01187 Dresden, Deutschland 
%\and 
%Signal Processing Lab IEETA, University of Aveiro, 3810-193 Aveiro,
%Portugal
\and Central Instrumentation,
Indian Institute of Chemical Biology (Council of Scientific and Industrial Research),
Kolkata 700032, India \and 
Potsdam Institute for Climate Impact Research, Telegraphenberg, Potsdam, Germany}
\date{Received: date / Revised version: date}
\abstract{We propose a rationale for experimentally studying the intricate
relationship between the rate of information transmission and synchronization
level in active networks, applying theoretical results recently proposed. We
consider two non-identical coupled Chua's circuit with non-identical coupling
strengths in order to illustrate the proceeding for experimental scenarios of
very few data points coming from highly non-coherent coupled systems, such
that phase synchronization can only be detected by methods that do not rely
explicitely on the calculation of the phase. A relevant finding is to show
that for the coupled Chua's circuit, the larger the level of synchronization
the larger the rate of information exchanged between both circuits. We further
validate our findings with data from numerical simulations, and discuss an
extension to arbitrarily large active networks.
%
%\PACS{
%      {05.45.-a}{ Nonlinear dynamics and chaos}   \and
%      {05.45.Xt}{ Synchronization; coupled oscillators}
%	{84.30.-r}{ Electronic circuits}
%     } % end of PACS codes
}%end of abstract
\maketitle
\section{Introduction}
\label{introduction}

Given an arbitrary time dependent stimulus that externally excites an active
network, formed by elements that have some intrinsic dynamics
(e.g. neurons or oscillators), how much information from such stimulus can be
realized by measuring the time evolution of one of the elements of the network
?  For example, in neurosciences, determining how and how much information
flows along anatomical brain paths is an important requirement for
understanding how animals perceive their environment, learn and behave
\cite{smith,eggermont,theunissen}.

Even though the approaches in
Refs. \cite{smith,eggermont,theunissen,roland,palus,dz} have brought
considerable understanding on how and how much information from a stimulus is
transmitted in a neural network, the relationship between synchronization and
information transmission in a neural as well as in an active network is still
awaiting a more quantitative description.

In order to treat this problem in a more analytical way, we proceed in the
same line as in Refs. \cite{schreiber,liang}, and study the information
transfer in autonomous networks. However, instead of treating the information
transfer between dynamical systems components, we treat the transfer of
information per unit time exchanged between two elements in an autonomous
chaotic active network \cite{baptista:2005} . Arguably, the relationship
between synchronization and information in autonomous chaotic networks is
useful for understanding its counterpart in non-autonomous active
networks.

The purpose of the present work is to revisit some previous theoretical
results and explain how to apply such approaches to study information
transmission and synchronization from data coming from experiments.

In Refs.
\cite{baptista:2005}, we proposed a formula [see Eq. (\ref{I_C})] that enables
the calculation of the rate with which information is exchanged between two
elements in a chaotic network, in terms of defined positive conditional
Lyapunov exponents.  Consider two non-identical coupled
chaotic systems with two positive conditional exponents, 
$\lambda^{\parallel}$ and $\lambda^{\perp}$. The upper bound for the
rate with which information is exchanged between these two elements is given
by $\lambda^{\parallel}-\lambda^{\perp}$.

While Lyapunov exponents measure the exponential divergence of nearby
trajectories in phase space, the conditional Lyapunov
exponents measure the exponential divergence of nearby trajectories on a
coordinate-transformed space. This transformed space (see Sec. \ref{MI}) is
constructed in such a way that if the elements in a network are almost
completely synchronous \cite{pecora,pecora1,comment}, then one conditional exponent,
$\lambda^{\parallel}$, measures the exponential divergence of trajectories
along the synchronization manifold, and the other exponent, $\lambda^{\perp}$,
measures the exponential divergence of trajectories along the transversal
manifold. Then, the rate of information exchanged between two elements is the
rate of information produced by the synchronous trajectories
($\lambda^{\parallel}$) minus the rate of information produced by the
desynchronous trajectories ($\lambda^{\perp}$). Thus, this formula enables one
to understand the relationship between information and synchronization, since
the so defined conditional exponents are a measure of the synchronization and
desynchronization between two elements in a network.

We apply the formula proposed in Refs.  \cite{baptista:2005} using an
experimental perspective. We consider that one has only a short time series
available to do the analysis and that the system is highly non-coherent. Under
such conditions, we will show that the largest Lyapunov exponent of a two
coupled Chua's circuit \cite{syamal} can only be well estimated using a
bivariate time series that contains information of the trajectories of both
circuits. Further, we show that the second largest Lyapunov exponent can only
be roughly estimated by using information from the characteristic of
conditional observations performed in one circuit while the other realizes
some event. These conditional observations, defined in
Refs. \cite{baptista_PHYSICAD2005,tiago:2007}, are in fact an alternative way
of detecting phase synchronization \cite{book_synchro,murilo_irrational}
without having to actually measure the phase. Such a method is a necessary
tool in order to study phase synchronization in non-coherent systems whose
phases might not always be well defined, as the one considered here, the two
non-identical diffusively coupled Chua's oscillators, with non-identical
coupling strengths (Sec. \ref{experiment_simulation}).

We start by showing how one can measure phase synchronization in this coupled
circuit (Sec. \ref{conditional_map}). Further, we demonstrate (Sec. \ref{MI})
that Eq. (\ref{I_C}) can be written in terms of the positive Lyapunov
exponents, thus enabling the use of standard codes to study information
transmission in coupled chaotic systems. Since the amount of data points in
each time series is small, alternative techniques to calculate the second
largest Lyapunov exponent will be developed (Sec. \ref{second_largest}). The
direct relationship between synchronization and information, one of the main
results of this work, is detailed in Sec. \ref{relacao}, and finally, in
Sec. \ref{extension}, we discuss how to extend our results to larger networks
with arbitrary connecting topologies.

\section{Experimental and numerical simulation setups}
\label{experiment_simulation}

\subsection{Experiment}

\begin{figure}
\centerline{\hbox{\psfig{file=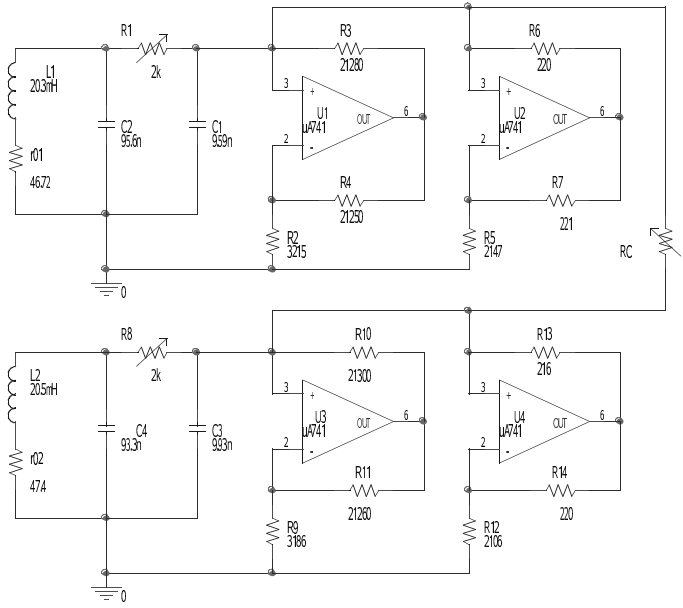,height=8.0cm,width=9cm}}}
%\resizebox{0.35\textwidth}{!}{\includegraphics{circuit.eps}}
\caption{Two diffusively coupled Chua's oscillators with a power supply of $\pm$9V,
  and parameters $L_1$=20.5$\mathrm{mH}$, $r_{01}$=47.46$\mathrm{\Omega}$,
  $C_1$=9.56$\mathrm{nF}$, $C_2$=95.9nF, $R_2$=3215$\mathrm{\Omega}$,
  $R_3$=21.28k$\mathrm{\Omega}$, $R_5$=2147$\mathrm{\Omega}$, $L_2$=20.3mH,
  $r_{02}$=46.98$\mathrm{\Omega}$, $C_3$=9.93nF, $C_4$=93.3nF,
  $R_9$=3186$\mathrm{\Omega}$, $R_{10}$=21.26k$\mathrm{\Omega}$,
  $R_{12}$=2106$\mathrm{\Omega}$.}
\label{chua_circuit}
\end{figure}

We consider two diffusively coupled non-identical Chua's circuits
\cite{syamal} as shown in Fig.~\ref{chua_circuit}.  Each oscillator is
composed by a resistor $R_{1,8}$, an inductor $L_{1,2}$, two capacitors
$C_{1,3}$ and $C_{2,4}$, and one piecewise-linear resistance. In our notation,
the first (second) index denotes an element in the upper (lower) circuit of
Fig. \ref{chua_circuit}.  The upper circuit is regarded as $S_1$ and the lower
as $S_2$. The piecewise-linear resistance is designed by a pair of linear
amplifiers, $U_1$-$U_2$ (in $S_1$) or $U_3$-$U_4$ (in $S_2$) with an op-amp $741$$\mathrm{\mu}A$,
for each oscillator. The resistance $R_C$ sets the coupling strengths, 
$\epsilon_1=\frac{R_1}{R_C}$ and $\epsilon_2=\frac{R_8}{R_C}$.

%\subsection{Data sets}

Two state variables, $x_1=V_{C1}$ and $x_2=V_{C3}$, are monitored using two
channels of a digital oscilloscope (Tektronix, TDS 220) at the nodes of the
capacitors $C_1$ and $C_3$, respectively, for varying coupling resistance
$R_C$. Data acquisition is made for 2500 data points at each snapshot by an
8-bit memory of the oscilloscope, with a time step-size $\Delta t$=0.002ms. All
circuit component values are precisely measured using a standard LCR-Q bridge
(APLAB 4910). We consider 20 data sets denoted by F$o$, with $o=\{1,\ldots,20\}$
representing the value of $R_C$. The larger $o$ is, the larger $R_C$
is. The set denoted as F21 contains data from the uncoupled circuits
($\epsilon$=0).

\subsection{Simulation}

To simulate the equations of motion of the circuit in Fig. \ref{chua_circuit},
we use the dimensionless set of equations given by 
%{\small
\begin{eqnarray}
\frac{dx_i}{d\tau} & = &
\tau_i \alpha_i[y_i-x_i-f(x_i)]+\epsilon_i \tau_i \alpha_i
(x_j-x_i) \nonumber \\ 
\frac{dy_i}{d\tau} & = & \tau_i(x_i-y_i+z_i)  \label{chuas} \\
\frac{dz_i}{d\tau} & = & \tau_i (-\beta_i y_i-\gamma_i z_i) \nonumber
\end{eqnarray}
\noindent
where $(i,j)=(1,2)$ with $j\ne i$, $\tau_1$=1, $\tau_2=\frac{R_1C_2}{R_8C_4}$,
$\alpha_1=\frac{C_2}{C_1}$, $\beta_1=\frac{R^2_1C_2}{L_1}$,
$\gamma_1=\frac{R_1r_{01}C_2}{L_1}$, $\alpha_2=\frac{C_4}{C_3}$,
$\beta_2=\frac{R^2_8C_4}{L_2}$, $\gamma_2=\frac{R_8r_{02}C_4}{L_2}$,
$\epsilon_1=\frac{R_1}{R_C}$, and $\epsilon_2=\frac{R_8}{R_C}$. The state
variables are the dimensionless voltages $x_{1}=\frac{V_{C1}}{E}$,
$x_{2}=\frac{V_{C3}}{E}$, $y_{1}=\frac{V_{C2}}{E}$, $y_{2}=\frac{V_{C4}}{E}$
(at the respective capacitor nodes), $z_{1}=\frac{R_{1}I_{L1}}{E}$, and
$z_{2}=\frac{R_{8}I_{L2}}{E}$ (where $I_{L1,L2}$ is the inductor current).
$E$ is the saturation voltage of the op-amps approximated as $E\approx 1$.

The parameters considered for the numerical simulations are
$r_{01}$=47.46, $r_{02}$=46.98, $R_1$=1650, $R_2$=3224, $R_3$=21300,
$R_4$=21330, $R_5$=2153, $R_6$=221.6, $R_7$=220.6, $R_8$=1650,
$R_9$=3194, $R_{10}$=21320, $R_{11}$=21330, $R_{12}$=2111,
$C_1$=9.56$\times$10$^{-9}$, $C_2$=95.9$\times$10$^{-9}$,
$C_3$=9.93$\times$10$^{-9}$, $C_4$=93.3$\times$10$^{-9}$,
$L_1$=20.5 $\times$ 10$^{-3}$ and
$L_2$=20.3 $\times$ 10$^{-3}$. Components have standard units as Ohm for
resistance, Farad for capacitance and Henry for inductance.

The piecewise-linear function $f(x_{1,2})$ is defined as
%{\small
\begin{equation}\label{eq:piecefun}
f(x_{1,2})=\left|\begin{array}{lllr} b_{1,2}x_{1,2}+(b_{1,2}-a_{1,2}), &
\mathrm{\ if} & x_{1,2}<-1 & \rightarrow \mathrm{\ \ \ Domain\ } D_- \\ a_{1,2}x_{1,2}, &
\mathrm{\ if} & -1 \leq x_{1,2} \leq
1 &  \rightarrow \mathrm{\ \ Domain\ }  D_0 \\ b_{1,2}x_{1,2}+(a_{1,2}-b_{1,2}), &
\mathrm{\ if} & x_{1,2}>1 &  \rightarrow \mathrm{\ \ \ Domain\ } D_+
\end{array} \right.
\end{equation}
%}
where $a_{1,2}=\big(-\frac{1}{R_{2,9}}-\frac{1}{R_{5,12}}\big)R_{1,8}$ and
$b_{1,2}=\big(\frac{1}{R_{3,10}}-\frac{1}{R_{5,12}}\big)R_{1,8}$.  The
piecewise linear function $f(x_{1,2})$ has a slope $a_{1,2}$ in the inner
region near the equilibrium at the origin (domain $D_0$) and a slope $b_{1,2}$
in the outer regions close to the two mirror symmetric equilibria of each
oscillator (domains $D_+$ and $D_-$).

The dimensionless variables in the time-$\tau$ frame of the numerical
simulations are obtained by rescaling the time-$t$ frame of the experiment by
$\tau = \frac{t}{R_1C_2}$.
%{\small
%\begin{equation}
%\tau = \frac{t}{R_1C_2}.
%\end{equation}
%}

\section{Phase, phase synchronization, and conditional maps}\label{conditional_map}

Phase synchronization (PS) \cite{book_synchro} is a phenomenon defined by
%{\small
\begin{equation}
|\Delta \phi(S_1,S_2)| = |\phi_1 - m \phi_2| \leq r,
\label{phase_synchronization}
\end{equation}
%}
\noindent
where $\phi_1$ and $\phi_2$ are the phases of two elements $S_1$ and $S_2$,
$m=\omega_2/\omega_1$ is the angular frequency ratio that can be a real number
\cite{murilo_irrational}, and $\omega_1$ and $\omega_2$ are the average
frequencies of oscillation of the elements $S_1$ and $S_2$. The phase $\phi$
is a function constructed on a 2D subspace, whose trajectory projection has
proper rotation, i.e. it rotates around a well defined center of rotation. The
Chua's circuit, while presenting a double scroll attractor, has no proper
rotation in the phase space, but it can have proper rotation in the
velocity space, therefore it can admit a phase that measures the
displacement of the tangent vector
\cite{baptista_PHYSICAD2005,tiago_PLA2007} and can be calculated as shown in
Ref. \cite{tiago_PLA2007} by 
%{\small
\begin{equation}
\phi(t)=\int_0^t \frac{\ddot{y}\dot{x}-\ddot{x}\dot{y}}{{(\dot{x}^2+\dot{y}^2)}}dt.
\label{phase_dxdy}
\end{equation}
%}

However, as neither the simulated nor the experimental circuit, for
$\epsilon\ne0$, present proper rotation in both phase and velocity spaces,
Eq. (\ref{phase_dxdy}) has only physical meaning for a time interval where the
attractors are far away from the equilibrium points, a time that can be large
but not infinitely large. Therefore, for the present study is necessary to
employ alternative methods that detect phase synchronization without having to
measure the phase, as the one proposed in
Refs. \cite{baptista_PHYSICAD2005,tiago:2007}.  If PS exists between two
subspaces, then by observing the trajectory of one circuit at the time the
other circuit makes a physical event (an event that has positive probability
of occurrence), there exists at least one special curve, $\Gamma$, in this
subspace, for which the points obtained from these conditional observations do
not visit its neighborhood. Such a curve $\Gamma$ is defined in the following
way. Given a point $x_0$ in the attractor projected onto the subspace of one
circuit where the phase is defined, $\Gamma$ is the union of all points for
which the phase, calculated from this initial point $x_0$, reaches $n
\langle r
\rangle$, with
$n=1,2,3,\ldots,\infty$ and $\langle r \rangle$ a constant (typically
2$\pi$). Clearly, an infinite number of curves $\Gamma$ can be defined.

For coupled systems with sufficiently close parameters that have proper
rotation in some
subspace, if the points obtained from the conditional
observations do not visit the whole attractor projection on this subspace, one
can always find a curve $\Gamma$ that is far away from the conditional
observations. Therefore, for such cases, to state the existence of PS one just
has to check if the conditional observations are localized with respect to the
attractor projection on the subspace where the phase is calculated.  Note 
that the value of the angular frequency ratio, $m$, is irrelevant to state PS using these
conditional mappings. Whatever $m$ is, if there is PS, these mappings will be
localized.

In a general situation, where the attractor has no proper rotation either in
phase or velocity spaces and the event is a physical event, thus, as demonstrated in
Ref. \cite{tiago:2007}, PS implies the localization of the conditional sets.

\subsection{Events}

An event is considered to be the crossing of the trajectory to 
a  Poincar\'e section. 

The experimental Poincar\'e sections are defined in the 2D time-delay space,
constructed using the coordinates ($x(t),x(t)+\delta$), with the time-delay
$\delta=6\Delta \tau$, and they are given by
%{\small
%\begin{equation}
%\left\{
\begin{eqnarray} 
x(t+\delta) &=& x_c \mathrm{\ and\ } x(t) \geq x_c \label{poincares_expI1} \\
x(t+\delta) &=& -x_c \mathrm{\ and\ } x(t) \leq -x_c \label{poincares_expI2}
\end{eqnarray}
%\right\}
%\end{equation}
%}
with $x_c$=1.5, for the data sets F1 to F16 plus F21, and
%{\small
\begin{equation}\label{poincares_expII}
%\left\{
\begin{array}{lllll} x(t+\delta)=x_c & & \mathrm{if} & &
x(t) \geq x_c\\
\end{array}
%\right\}
\end{equation}
%}
with $x_c$=0, for the data sets F17 to F20. The theoretical Poincar\'e
sections are defined as
%\begin{equation}
%\left\{
\begin{eqnarray} 
x_1(t)&=&x_c \mathrm{\ and\ } y_1(t) < 0  \label{poincares_sim1} \\
x_1(t)&=&-x_c \mathrm{\ and\ } y_1(t) < 0 \label{poincares_sim2}
\end{eqnarray}
%\right\}
%\end{equation}
\noindent
with $x_c=2$.

\subsection{Observing phase synchronization in the coupled Chua's circuit without measuring the phase} 

\begin{figure}
\centerline{\hbox{\psfig{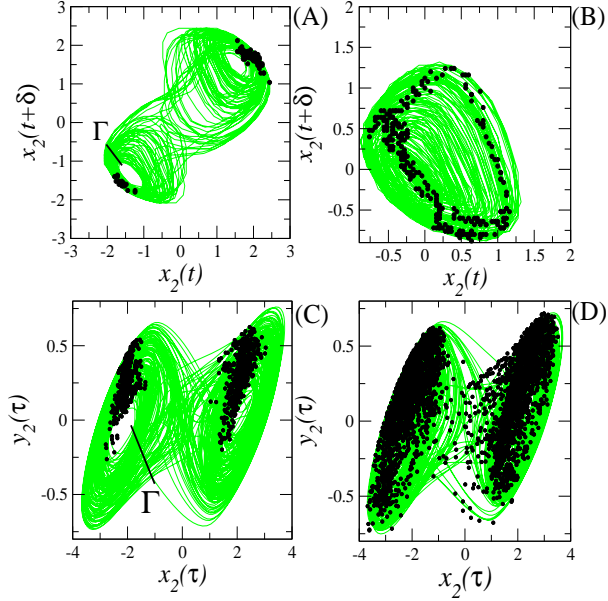}}}
%\resizebox{0.45\textwidth}{!}{\includegraphics{syamal_02.eps}}
\caption{[Color online] Projections of the attractor [gray (green) lines] 
and conditional mappings [black filled circles]. Experimental results are
shown in (A-B) and simulations in (C-D). PS happens for the data set F12 (A)
and it is absent for the data set F18 (B). PS is observed for
$R_C$=12,000 (C) and is absent for $R_C$=25,000 (D). The conditional mappings
for (A) [resp. (B)] are constructed by observing the circuit $S_2$ at the
moment the events defined in conditions (\ref{poincares_expI1}) and
(\ref{poincares_expI2}) [resp. Eq. (\ref{poincares_expII})] happen in $S_1$,
and the conditional mappings in (C-D) are constructed by observing the circuit
$S_2$ at the moment the events defined in conditions (\ref{poincares_sim1}) and
(\ref{poincares_sim2}) happen in $S_1$. The straight black line (A,C) illustrates a
surface $\Gamma$.}
\label{syamal_02}
\end{figure}

In Figs. \ref{syamal_02}(A,C), we show the presence of PS in the experiment and
in the simulations, respectively, while in Figs. \ref{syamal_02}(B,D), we show
the absence of such a phenomenon. While in Figs. (A,C), a surface $\Gamma$ can
be defined such that the conditional observations do not visit it, i.e. the
conditional observations are localized with respect to the attractor, in
(B,D) the conditional observations spread all over the attractor, i.e. they
are not localized.

\section{Mutual information rate, Lyapunov and conditional exponents}
\label{MI}

In recent publications \cite{baptista:2005}, we have shown that the mutual
information rate (MIR) between two elements in an active chaotic network,
quantifying the amount of information per unit time that can be realized in one
element, $i$, by measuring another element, $j$, is given
by the sum of the conditional Lyapunov exponents associated with a parallel
coordinate transformation minus the positive conditional Lyapunov exponents
associated with a transversal coordinate transformation.

Assuming that every element possesses only one positive Lyapunov exponent, for
every pair of elements, whose state variables are given by ${\bf x}_i$ and
${\bf{x}}_j$, we can define a coordinate transformation ${\bf
x}^{\parallel}_{ij}={\bf{x}}_i+{\bf{x}}_j$ and ${\bf x}^{\perp}_{ij}={\bf x}_i
- {\bf x}_j$ that produces two positive conditional exponents, 
$\lambda^{\parallel}$ and $\lambda^{\perp}$ (in units of
bits/unit time). The mutual
information rate (MIR), denoted by $I_C(t)$, between the element ${\bf x}_i$
and ${\bf{x}}_j$ is bounded from above by $\lambda^{\parallel}-\lambda^{\perp}$,
and thus
%{\small
\begin{equation}
I_C({\bf x}_i,{\bf{x}}_j) \leq \lambda^{\parallel}-\lambda^{\perp}
\label{I_C}
\end{equation}
%}
\noindent
where equality certainly holds if the elements are identical and are either in
complete synchrony or decoupled ($\epsilon=0$).

As shown in Ref. \cite{baptista:2005}, if there are N=2 linearly coupled chaotic
systems that produce at most two positive Lyapunov exponents, $\lambda^1$ and
$\lambda^2$, with $\lambda^1>\lambda^2$, then $\lambda^{\parallel} =
\lambda^1$ and $\lambda^{\perp}=\lambda^2$, since the parallel and the
transversal coordinate transformations are only rotations which do not alter
the value of the Lyapunov exponents.

%Whenever necessary, we will identify the Lyapunov and conditional exponents
%obtained from the experiment (simulation) by the introduction of an lower
%index $e$. So, the positive Lyapunov and conditional exponents obtained from
%the experimental data are denoted by $\lambda^1_e, \lambda^2_e$ and
%$\lambda^{\parallel}_e, \lambda^{\perp}$, respectively.

This result can be easily demonstrated for the system considered here, due to
its linear form. Thus, we can write
%{\small
\begin{equation}
I_C({{\bf x}}_i,{\bf{x}}_j) \leq \lambda^{1}-\lambda^{2}
\label{I_C2}
\end{equation}
%}
\noindent
Making the notation ${\bf{x}} = ({\bf{x}}_1^T,{\bf{x}}_2^T)$ and ${\bf{X}} =
({{\bf{x}}^{\perp T}_{12},{\bf{x}}^{\parallel T}_{12}})$, we have that
%{\small
\begin{eqnarray}
{\bf{\dot{x}}} &=& {\bf{M}_1} {\bf{x}} + c_1 \label{m1} \\
{\bf{\dot{X}}} &=& {\bf{M}_2} {\bf{X}} + c_2 \label{m2} \\ 
{\bf{X}} &=& {\bf{M}} {\bf{x}} \label{m}
\end{eqnarray}
%}
where $\bf{M}_1$, $\bf{M}_2$, and $\bf{M}$ are 6$\times$6 matrices and $c_1$
and $c_2$ are constant terms from the piecewise-linear function.
Matrices $\bf{M}_2$ and $\bf{M}$ are explicitly written in Appendix
(Sec. \ref{apendiceA}), while  matrix
$\bf{M}_1$ is the Jacobian of Eqs. (\ref{chuas}).

Writing Eqs. (\ref{m1}), (\ref{m2}), and (\ref{m}) in the variational form,
and making a Taylor expansion (which eliminates the constant terms), the
following equations are retrieved
%{\small
\begin{eqnarray}
{\bf{\dot{\xi x}}} &=& \bf{M}_1   {\bf{\xi x}} \label{m1_d}, \\
{\bf{\dot{\xi X}}} &=& \bf{M}_2   {\bf{\xi X}} \label{m2_d}.
\end{eqnarray}
%}
While the Lyapunov exponents of Eqs. (\ref{chuas}) are calculated from
Eq. (\ref{m1_d}), the conditional exponents are calculated from
Eq. (\ref{m2_d}), both using the approach in Ref. \cite{ruelle}. But,
%{\small
\begin{equation}
{\bf{\dot{\xi x}}} = {\bf{M}}^{-1}.{\bf{M}}_2.{\bf{M}} {\bf{\xi x}} \label{m3_d}. 
\label{transforma}
\end{equation}
%}
\noindent
Noting that ${\bf{M}}^{-1}.{\bf{M}}_2.{\bf{M}}$ is just a rotation applied to matrix
$\bf{M}_2$, and since a rotation does not change the eigenvalues of $\bf{M}_2$, thus, the
Lyapunov exponents should be equal to the conditional exponents.

Assuming that we have a large active network, the theoretical approaches proposed in
\cite{baptista:2005} remain valid whenever the coordinate transformation
${\bf{x}}_{ij}^\parallel={\bf{x}}_{i}+{\bf{x}}_{j}$ and ${\bf{x}}_{ij}^\perp=
{\bf{x}}_{i}-{\bf{x}}_{j}$ successfully separate the two systems $i$ and
$j$ from the whole network. Such a situation arises, for example, in networks of chaotic
maps of the unit interval connected by a diffusive (also known as electrical or
linear) all-to-all topology, where every element is connected to all other
elements. These approaches were also shown to be approximately valid for
chaotic networks of oscillators connected by a diffusively all-to-all
topology. The discussion on how to extend such approaches to arbitrary network
topologies is given in Sec. \ref{extension}.

In order to compare our results with known quantities, we will also calculate
the MIR using Shannon's formalism \cite{shannon}.  The MIR between the two circuits
can be roughly estimated by symbolizing their trajectories and then measuring
the mutual information from the Shannon entropy of the symbolic
sequences.  The mutual information between $S_1$ and
$S_2$ is given by
%{\small
\begin{equation}
I_{S}^{\prime}=H(S_1)-H(S_2|S_1), 
\label{shannon1}
\end{equation}
%}
\noindent
where $H(S_1)$ is the uncertainty about what $S_1$ has sent (entropy of the
message), and $H(S_2|S_1)$ is the uncertainty of what was sent, after
observing $S_2$.  In order to estimate the mutual information between the two
chaotic Chua's circuit by symbolic ways, we have to proceed with a non-trivial
technique to encode the trajectory, which constitutes a disadvantage of such
technique to chaotic systems.  We represent the time at which the $n$-th event
happens in $S_k$ ($k$=\{1,2\}) by $T_k^n$, and the time interval between the
$n$-th and the ($n$+1)-th event, by $\delta T_k^n$.  

We encode the events using the following rule. The $i$-th symbol of the
encoding is a ``1'' if an event is found in the time interval $[i\Delta, (i+1)
\Delta[$ and ``0'' otherwise.  We choose $\Delta \in [\min{(\delta T_k^n)},
\max{(\delta T_k^n)}]$ in order to maximize $I_S^{\prime}$. Each circuit produces a
symbolic sequence that is split into small non-overlapping sequences of length
$l$=12. The Shannon entropy of the encoding symbolic sequence (in units of
bits) is estimated by $H$ = -$\sum_p P_p \log_2 P_p$ where $P_p$ is
the probability of finding one of the 2$^l$ possible symbolic sequences of
length $l$. The term $H(S_2|S_1)$ is calculated by
$H(S_2|S_1)$=$-H(S_2)+H(S_1;S_2)$, with $H(S_1;S_2)$ representing the joint
entropy between both symbolic sequences for $S_1$ and $S_2$.

Finally, the MIR (in units of bits/unit time), denoted by $I_S$, is calculated from 
%{\small
\begin{equation}
I_S = \frac{I_S^{\prime}}{\Delta \times l}.
\label{shannon}
\end{equation}
%}

The calculation of $I_S$ by means of Eq. (\ref{shannon}) should be
expected to underestimate the real value of the MIR. Since the Chua's circuit
has two time-scales, a large sequence of sequential zeros in the encoding
symbolic sequence should be expected to be found between two events (large
$\delta T_k^n$ values), leading to a reduction in the value of $H(S_1)$, 
followed by an increase in the value of $H(S_2|S_1)$, as there will be a
large sequence of zeros happening simultaneously in the encoding sequence for
the time intervals between two events of $S_1$ and $S_2$.

\subsection{Experimental exponents}
\label{second_largest}

The estimation of the Lyapunov exponents from the experimental time series
data was done using a method recently proposed (Ref. \cite{sara_ref4}). The
first step in the algorithm is the phase reconstruction, accomplished by means of the nearest neighbor
embedding with different time delays method proposed in Refs.
\cite{sara_ref1}. This method considers different time delays for every
embedding coordinate. The embedding dimension is estimated using the false
nearest neighbors criterion proposed in Ref. \cite{kennel}. The second step of
this algorithm pertains estimating local tangent maps by a least-squares
minimization with a pseudo-inverse method. Finally, in the third step of this
method, the exponents are derived from the usual QR decomposition with a
modified Gram-Schmidt method.

Due to the small number of data points and the additional fact that the
coupled Chua's circuit has a highly non-coherent dynamics, a better estimate
of the largest Lyapunov exponents was achieved by an attractor reconstructed
from the bivariate data set ($x_1(t),x_2(t)$).  

However, even the bivariate data set is not capable of providing a second
largest positive Lyapunov exponent, $\lambda^2$, which should be positive if
there is not complete synchronization. So, in order to estimate $\lambda^2$
from the experimental data sets, we assume that
%{\small
\begin{equation}
\lambda^2=\lambda^1 \left(\frac{\max{(x_2^n)} -
\min{(x_2^n)}}{\max{(x_2)}-\min{(x_2)}} \right), 
\label{estima_lambda2}
\end{equation}
%}
\noindent
where $x_2^n$ is the value of $x_2(t+\delta)$ at the moment the circuit $S_1$
makes its $n$-th event. By an event, we consider
conditions (\ref{poincares_expI1}) and (\ref{poincares_expII}).  While
[$\max{(x_2^n)}-\min{(x_2^n)}$] measures the size of the conditional
observations, [$\max{(x_2)}-\min{(x_2)}$] measures the size of the
reconstructed attractor. 

\begin{figure}
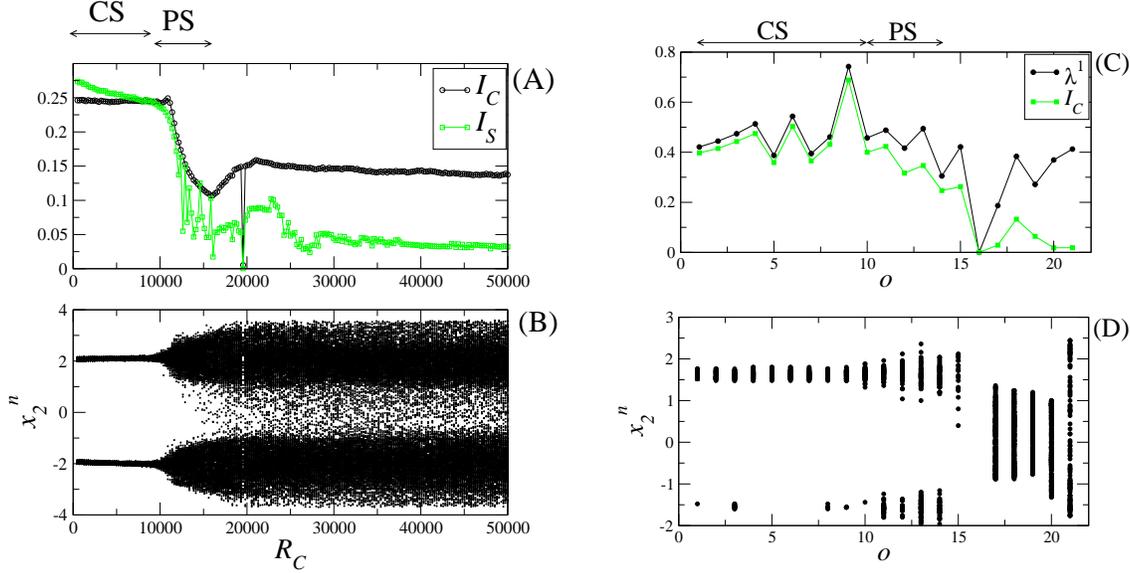

\centerline{\hbox{\psfig{file=syamal_01.eps,width=8cm,viewport=0 -20 500 500}}
\hfill  \hbox{\psfig{file=syamal_04.eps,width=7.5cm,viewport=0 -30 500 500}}}
%\resizebox{0.4\textwidth}{!}{\includegraphics{syamal_01.eps}}
\caption{[Color online] Simulations are shown in (A-B) and experimental
results in (C-D). In (A), we show $I_C$ [see Eq. (\ref{I_C2})] and $I_S$ [see
Eq. (\ref{shannon})]. The Lyapunov exponents of the simulated circuit, with
which $I_C$ is calculated, are obtained using the method of Ref. 
\cite{ruelle} and the variational equations in Eq. (\ref{m1_d}).
Complete synchronization (CS) in the generalized sense \cite{pecora1,comment} is
observed for $R_C<9000$ and PS for $9000 \leq R_C \leq 16000$. In (B), 
we show the conditional observations realized in $S_2$ at the moment $S_1$ makes its
$n$-th event, i.e. the crossing of $x_1(\tau)$ with the defined Poincar\'e
sections [conditions (\ref{poincares_sim1}) and (\ref{poincares_sim2})].  A
periodic orbit is observed for $R_C \cong 20,000$. In (C), we show $\lambda^1$
calculated as described in Sec. \ref{second_largest}, and $I_C$ is calculated 
considering that $\lambda^2$ is estimated from
Eq. (\ref{estima_lambda2}). CS in the generalized sense
is observed for the data series F$o$, with $o\leq10$, and PS for F$o$, with $
11 \leq o\leq 14$. In (D), we show the conditional observations realized in
$S_2$ at the moment $S_1$ makes its $n$-th event, i.e. the crossing of
$x_1(t)$ with the defined Poincar\'e sections [see
conditions (\ref{poincares_expI1}), (\ref{poincares_expI2}), and
(\ref{poincares_expII})]. A periodic orbit is observed for the data set
$F16$.}
\label{syamal_01}
\end{figure}

Thus, if the conditional observations cover the whole attractor, 
$\lambda^2=\lambda^1$,  and thus $I_C$=0, which means no information is being
transmitted between both circuits, since whenever $S_1$ crosses the defined
Poincar\'e section, $S_2$ can be everywhere. If there is complete
synchronization in the generalized sense \cite{pecora1,comment}, [$\max{(x_2^n)} -
\min{(x_2^n)}] \cong$0 and $I_C$=$\lambda^1$, meaning that whenever $S_1$
crosses the defined Poincar\'e section, $S_2$ is also about to or has just
crossed that particular section.  Therefore, the information about one circuit trajectory by
observing the other circuit is maximal.

\section{Synchronization versus information}\label{relacao}

In Figs. \ref{syamal_01}(A-B), we show results from our numerical simulations,
while in Figs. (C-D), experimental results. In both cases, one obvious
observation is that the more synchronous the circuits are (small $R_C$), the larger the
rate of information that can be measured in one circuit about the other
circuit,  being maximal when the two circuits are completely
synchronous in the generalized sense \cite{pecora1,comment}. When both circuits
are in PS, the MIR decreases but remains larger than when there is no PS.

\section{Mutual information rate in large active networks}\label{extension}

For large active networks with elements arbitrarily connected, an extension of Eq. (\ref{I_C}) is
\begin{equation}
I_C({\bf x}_i,{\bf{x}}_j) \leq \max{(\lambda)} - \lambda^{\perp}, 
\label{I_Clarge}
\end{equation}
\noindent
where $\max{(\lambda)}$ is the largest Lyapunov exponent of the network and
$\lambda^{\perp}$ is the transversal exponent between the elements ${\bf x}_i$
and ${\bf{x}}_j$. Making an analogy with the usual
definition of mutual information as given by Shannon \cite{shannon}, the term
$\max{(\lambda)}$ provides the rate of information produced by the source, 
and the term $\lambda^{\perp}$ quantifies the error in the transmission. The term
$\max{(\lambda)}$ can be calculated by a scalar signal measured from ${x}_i$, 
and $\lambda^{\perp}$ can be estimated either by the ways of
Eq. (\ref{estima_lambda2}) or as similarly done in Ref. \cite{tiago:2007}.

\section{Conclusions}\label{conclusions}

This work proposes a rationale to experimentally study the relationship between
transmission of information and synchronization in active networks formed by non-identical
and non-coherent elements. For the coupled Chua's circuit,
we have shown that the larger the level of synchronization the larger 
the rate of information exchanged between both circuits, which implies that such
a system is non-excitable.  By non-excitability \cite{baptista:2005} we mean
a system that as the coupling strength increases, the Kolmogorov-Sinai
entropy
\cite{ruelle} (the sum of the positive Lyapunov exponents) decreases. For such systems,
the maximal mutual information rate that can be achieved, the so called
channel capacity, happens for when complete (generalized) synchronization is
present.

Other relevant contributions of this work include showing that for short
time series the largest conditional exponent (demonstrated to be identical to
the largest Lyapunov exponent) can only be reliably estimated by using a
multivariate data set, with information of both elements being
considered, and that the second largest conditional exponent (which also equals the
second largest Lyapunov exponent) can only be reliably estimated by the conditional
observations, realized in one element when the other
makes an event. Finally, we have also shown that for two coupled
non-identical and non-coherent systems, phase synchronization can be detected
by these conditional observations, even though for such a system phase is not
well defined.

\section{Appendix}\label{apendiceA}

Consider the notation ${\bf{x}} = ({\bf{x}}_1^T,{\bf{x}}_2^T)$ and ${\bf{X}} =
({\bf{x}}^{\perp T}_{12},{\bf{x}}^{\parallel T}_{12})$, such that 
{\small
\begin{eqnarray}
{\bf{\dot{x}}} &=& \bf{M} {\bf{x}} \mathrm{\ \ , and}\nonumber \\
{\bf{\dot{X}}} &=& \bf{M}_2 {\bf{X}} + c_2 \mathrm{\ \ ,} \nonumber
\end{eqnarray}}
\noindent
where

{
%\small
\parbox{6cm}{
%\begin{equation}
$$
%\label{matriz M}
{\bf{M}}=\left(\begin{array}{rrrrrr} 1 & 0 & 0 & -1 & 0 & 0\\ 0 & 1 & 0 &
0 & -1 & 0\\ 0 & 0 & 1 & 0 & 0 & -1\\ 1 & 0 & 0 & 1 & 0 & 0\\ 0 &
1
& 0 & 0 & 1 & 0\\ 0 & 0 & 1 & 0 & 0 & 1\\
\end{array}\right)
%\end{equation}
$$
}
}
%\noindent
%and
\parbox{6cm}{
{\Large
\begin{equation}\label{jacobiano}
{\bf{M}}_2=\left(\begin{array}{lll}
\frac{\partial\dot{x}_{12}^\perp}{\partial x_{12}^\perp} & &
\frac{\partial\dot{x}_{12}^\perp}{\partial x_{12}^\parallel}, \nonumber \\
\frac{\partial\dot{x}_{12}^\parallel}{\partial x_{12}^\perp} & &
\frac{\partial\dot{x}_{12}^\parallel}{\partial x_{12}^\parallel} \nonumber \\
\end{array}\right)
\end{equation}}
}

\noindent
with the terms in matrix (\ref{jacobiano}) given  by

\parbox{8cm}{
{\small
$
\frac{\partial \dot{x}_{12}^\perp}{\partial x_{12}^\perp} =
\left(\begin{array}{rrrrr} \left(-\sigma_1
-\frac{\partial g_1(x_{12}^\perp,x_{12}^\parallel)}{\partial x^\perp}
-\sigma_2\right) & & \sigma_1 & & 0\\
\sigma_3 & & -\sigma_3 & & \sigma_3\\ 0 & & -\sigma_4 & & -\sigma_5\\
\end{array}\right), 
$}}
\hfill
\parbox{8.0cm}{
{\small
$
\frac{\partial \dot{x}_{12}^\perp}{\partial x_{12}^\parallel} =
\left(\begin{array}{rrrrr} \left(-\sigma_8
-\frac{\partial g_1(x_{12}^\perp,x_{12}^\parallel)}{\partial x^\parallel}\right) & & \sigma_8 & & 0\\
\sigma_7 & & -\sigma_7 & & \sigma_7\\ 0 & & -\sigma_{9} & & -\sigma_{10}\\
\end{array}\right), 
$}}

\parbox{8.0cm}{
{\small
$
\frac{\partial \dot{x}_{12}^\parallel}{\partial x_{12}^\perp} =
\left(\begin{array}{rrrrr} \left(-\sigma_8
-\frac{\partial g_2(x_{12}^\perp,x_{12}^\parallel)}{\partial x^\perp}+\sigma_6\right) & & \sigma_8 & & 0\\
\sigma_7 & & -\sigma_7 & & \sigma_7\\ 0 & & -\sigma_{9} & & -\sigma_{10}\\
\end{array}\right), 
$}}
\hfill
\parbox{8.0cm}{
{\small
$
\frac{\partial \dot{x}_{12}^\parallel}{\partial x_{12}^\parallel} =
\left(\begin{array}{rrrrr} \left(-\sigma_1
-\frac{\partial g_2(x_{12}^\perp,x_{12}^\parallel)}{\partial x^\parallel}\right) & & \sigma_1 & & 0\\
\sigma_3 & & -\sigma_3 & & \sigma_3\\ 0 & & -\sigma_4 & & -\sigma_5\\
\end{array}\right), 
$}}
\noindent
where 
$\sigma_1=\frac{\alpha_1+\alpha_2\tau}{2}$,  
$\sigma_2=\Big(\frac{\alpha_1R_1}{R_C}+\frac{\alpha_2\tau
R_8}{R_C}\Big)$, $\sigma_3=\frac{1+\tau_C}{2}$, $\sigma_4=\frac{\beta_1+\beta_2\tau}{2}$,
$\sigma_5=\frac{\gamma_1+\gamma_2\tau}{2}$, 
$\sigma_6=\Big(\frac{\alpha_2\tau
R_8}{R_C}-\frac{\alpha_1R_1}{R_C}\Big)$, 
$\sigma_7=\frac{1-\tau_C}{2}$, $\sigma_8=\frac{\alpha_1-\alpha_2\tau}{2}$,
$\sigma_{9}=\frac{\beta_1-\beta_2\tau}{2}$,
$\sigma_{10}=\frac{\gamma_1-\gamma_2\tau}{2}$.

The terms $\frac{\partial g_1(x_{12}^\perp,x_{12}^\parallel)}{\partial
x^\perp}$, $\frac{\partial g_1(x_{12}^\perp,x_{12}^\parallel)}{\partial
x^\parallel}$, $\frac{\partial g_2(x_{12}^\perp, x_{12}^\parallel)}{\partial
x^\perp}$, and $\frac{\partial g_2(x_{12}^\perp, x_{12}^\parallel)}{\partial
x^\parallel}$ assume different values depending on which of the domains,
namely (i), (ii), (iii), (iv), the values of $x_1$ and $x_2$ belong to, and are
given by

\parbox{8.0cm}{
\begin{tabular}{ccccc}
\hline\noalign{\smallskip}
Domains & (i) & (ii) & (iii) & (iv)\\
\noalign{\smallskip}\hline\noalign{\smallskip}
$\frac{\partial g_1(x_{12}^\perp,x_{12}^\parallel)}{\partial x_{12}^\perp}$ & $\xi_8$ & $\xi_5$ & $\xi_1$ & $\xi_4$ \\
$\frac{\partial g_1(x_{12}^\perp,x_{12}^\parallel)}{\partial x_{12}^\parallel}$ & $\xi_7$ & $\xi_6$ & $\xi_2$ & $\xi_3$ \\
$\frac{\partial g_2(x_{12}^\perp,x_{12}^\parallel)}{\partial x_{12}^\perp}$ & $\xi_7$ & $\xi_6$ & $\xi_2$ & $\xi_3$ \\
$\frac{\partial g_2(x_{12}^\perp,x_{12}^\parallel)}{\partial x_{12}^\parallel}$ & $\xi_8$ & $\xi_5$ & $\xi_1$ & $\xi_4$ \\
\noalign{\smallskip}\hline
\end{tabular}}
\hfill
\parbox{8.0cm}{\begin{tabular}{c}
$\xi_1=\frac{a_1\alpha_1+b_2\alpha_2\tau}{2}$ \\
$\xi_2=\frac{a_1\alpha_1-b_2\alpha_2\tau}{2}$ \\
$\xi_3=\frac{a_1\alpha_1-a_2\alpha_2\tau}{2}$ \\
$\xi_4=\frac{a_1\alpha_1+a_2\alpha_2\tau}{2}$ \\
$\xi_5=\frac{b_1\alpha_1+a_2\alpha_2\tau}{2}$ \\
$\xi_6=\frac{b_1\alpha_1-a_2\alpha_2\tau}{2}$ \\
$\xi_7=\frac{b_1\alpha_1-b_2\alpha_2\tau}{2}$\\
$\xi_8=\frac{b_1\alpha_1+b_2\alpha_2\tau}{2}$\\
\end{tabular}}
\noindent
where domain (i) is defined by $x_1 \in D_-$ and $x_2 \in D_-$, or $x_1
\in D_-$ and $x_2 \in D_+$; domain (ii) by $x_1 \in D_-$ and $x_2 \in D_0$;
domain (iii) by $x_1 \in D_0$ and $x_2 \in D_-$ or $x_2 \in D_+$;  
and domain (iv) by $x_1 \in D_0$ and $x_2 \in D_0$.

\end{document}